# Interpretable Cyber Threat Detection for Enterprise Industrial Networks: A Computational Design Science Approach

*Completed Research Paper*

**Prabhat Kumar**
Department of Software Engineering, LUT University, 53850 Lappeenranta, Finland
prabhat.kumar@lut.fi

**A.K.M. Najmul Islam**
Department of Software Engineering, LUT University, 53850 Lappeenranta, Finland
najmul.islam@lut.fi

**Abstract**

Enterprise industrial networks face threats that risk data and operations. However, designing efficient threat detection system is challenging due to data scarcity, especially where privacy is a concern. The complexity of enterprise industrial network data adds to this challenge, causing high false positives and interpretation issues. Towards this, we use IS computational design science paradigm to develop a two-stage cyber threat detection system for enterprise-level IS that are both secure and capable of adapting to evolving technological and business environments. The first stage generates synthetic industrial network data using a modified generative adversarial network. The second stage develops a novel bidirectional gated recurrent unit and a modified attention mechanism for effective threat detection. We also use shapley additive explanations and a decision tree technique for enhancing interpretability. Our analysis on two public datasets shows the framework's high precision in threat detection and offers practical cybersecurity solutions and methodological advancements.

## Introduction

In today's hyper-connected world, organizational systems are crucial connections for the flow and analysis of information across various sectors, particularly within industrial networks (Beverungen et al. 2021). In this ecosystem, enterprise systems (ES) play a crucial role in controlling and automating essential business activities across a range of industries, including industrial operations, which are increasingly reliant on interconnected technologies (Yu Chung Wang et al. 2022). These systems gather, analyze, and oversee data across different functions, making them beneficial in implementing robust Information Systems (IS) management strategies. These strategies are pivotal for integrating advanced technologies such as Industrial Internet of Things (IIoT), machine learning algorithms, and real-time data processing, which enhance operational efficiencies, proactive maintenance, and process optimizations within enterprise industrial networks (Unhelkar and Arntzen 2020). The importance of these technologies is underscored by the rapid expansion of the industrial networking solutions market, projected to grow from USD 29.05 billion in 2023 to USD 150.68 billion by 2032 at a CAGR of 20.1% (Research 2024). This surge in market growth highlights the increasing reliance on digital infrastructures and underscores the critical need for robust cybersecurity measures.

Cybersecurity has become a fundamental pillar for safeguarding enterprise industrial networks, which are increasingly exposed to sophisticated cyber threats due to integrating IIoT and other digital technologies. These attacks can disrupt critical operations, steal sensitive data, and endanger intellectual property, underscoring the urgent need for robust cyber threat detection systems (Singh et al. 2023). The importance of these systems in such an environment is twofold: they not only help to identify and mitigate potential threats before they cause harm proactively, but they also play a crucial role in ensuring operational continuity and safeguarding business assets. Traditional methods for cyber threat detection, such as rulebased systems and signature matching, often fall short due to their static nature and inability to adapt to new threats. In contrast, deep learning-based models offer significant advantages due to their ability to learn complex patterns and anomalies from data, adapt over time, and detect previously unseen types of attacks (Kumar et al. 2022). However, developing effective deep learning models for cyber threat detection in enterprise industrial networks faces several challenges. Primarily, the scarcity of comprehensive and diverse datasets due to strict privacy regulations significantly hampers the training and effectiveness of these models (De et al. 2022). Unlike other domains where synthetic data generation is more established, cybersecurity—particularly within industrial networks—poses unique challenges due to the need to accurately replicate complex, dynamic attack patterns and network behaviors that are critical for effective threat detection. Moreover, the complex and sequential nature of network traffic data often leads to high rates of false positives, complicating the reliable detection of genuine threats (Nedeljkovic and Jakovljevic 2022). Finally, the opaque nature of deep learning models poses a significant barrier to their acceptance and trustworthiness, as stakeholders require comprehensible insights into the decision-making processes to validate and trust the outcomes produced by these models (Johnson et al. 2022). These limitations necessitate a novel Information Technology (IT) artifact capable of generating synthetic data, identifying cyber threats, and understanding the decision-making process in enterprise industrial networks.

Developing new IT artifacts (e.g., systems, methods, and models) to solve practical problems is a key part of of IS research (Hevner et al. 2008). This area of design research is becoming more important with the rise in interest in the IIoT and predictive analytics. Our research aims to gain useful insights by creating advanced analytics methods (e.g., deep learning) by using publicly available new data sources that were previously unavailable or not fully used (e.g., data from industrial sensors). Towards this, we adopt the computational design science paradigm (Yu et al. 2023) to develop and evaluate a novel cyber threat detection framework with two stages: (1) a novel Modified Conditional Generative Adversarial Network (MCGAN) to generate synthetic network traffic data that reflects the sequential and complex nature of real-world enterprise industrial network traffic. (2) a novel neural network architecture that integrates Bidirectional Gated Recurrent Units (BiGRU) and a Modified Bahdanau Attention Mechanism (MBAM) to process the synthetic data and make informed decisions. In this architecture, the BiGRU component effectively processes the time-series data in both forward and backward directions to capture both past and future context from the industrial data. This is further used by the MBAM component, which dynamically weighs the importance of different features in the input data, allowing the model to focus on the most relevant features for decision-making. Furthermore, the interpretation of proposed model is achieved using SHapley Additive exPlanations (SHAP) and Decision Tree Surrogate (DTS) technique. Consistent with the guidelines of the design science (Hevner et al. 2008) and computational design science paradigm (Ampel et al. 2024), we evaluated our proposed cyber threat detection framework with a series of benchmark experiments using two publicly available industrial datasets namely: ToN-IoT (Alsaedi et al. 2020) and Edge-IIoT (Ferrag et al. 2022).

The remainder of this paper is organized as follows. First, we review recent IS literature on synthetic data generation and cyber threat detection using various computational models. Second, we identify research gaps in the reviewed literature and pose our study's research questions. Third, we present the major components of our research design. Subsequently, we present the results, contributions to the IS knowledge base, and practical implications of this study. Finally, we conclude this research and suggest promising future research directions.

## Literature Review

This section reviews recent IS literature on synthetic data generation and cyber threat detection using various computational models. This examination highlights the prior methodologies and identifies significant research gaps that guide this research.

## Computational Models for Synthetic Data Generation

The limited availability of comprehensive data in enterprise contexts presents significant challenges, particularly where privacy concerns restrict the use of sensitive information. This issue is well-documented in the literature (Quach et al. 2022), emphasizing the challenge between the need for detailed data to drive business insights and the imperative to safeguard individual privacy (Kisselburgh and Beever 2022). Towards this, various research articles have used Generative Adversarial Network (GAN) for synthetic data generation. For example, Lin et al. (2022) proposes a framework named "IDSGAN" to generate adversarial malicious traffic records of attack and test the robustness of intrusion detection systems. Similarly, Mozo et al. (2022), Anande and Leeson (2023) used GAN to generate synthetic network data that can closely mimic the attack behaviours. We can see that GANs are effective in generating diverse attack scenarios but are limited to the authenticity and variability of the generated data. It is important that the synthetic data should reflect the complex real-world attack patterns without any biases (Lim et al. 2024). Moreover, traditional GANs often requires domain-specific knowledge to generate contextually accurate and useful data. This frequently necessitates extensive domain knowledge and further model tuning to ensure that synthetic data generation matches specific security scenarios. Building on the foundation of GANs, Yoon et al. (2019) proposed TimeGAN, a novel generative model designed to produce realistic time-series data while preserving temporal dynamics. By integrating both supervised and adversarial training objectives, TimeGAN ensures that the synthetic data maintain the inherent time-dependent relationships observed in real datasets. Yang et al. (2023) proposed Long Short-Term Memory (LSTM) networks based GANs named TSGAN for creating realistic data in healthcare. By employing LSTM networks, TS-GAN effectively captures and replicates the temporal dependencies inherent in physiological data, making the synthetic data it generates highly representative of real patient metrics. Furthermore, to improve the model ability to identify and highlight important features they integrated sequential-squeeze-and-excitation (SSE) module.

## Computational Models for Cyber Threat Detection

Recently, IS scholars and the broader Computer Science (CS) community has given significant attention to deep learning techniques for the development of cyber threat detection systems. For instance, Kayhan et al. (2023) designed an unsupervised framework for hunting anomalous commands that can identify unusual command activities within security information and event management logs. The framework used an autoencoder to detect anomalies. Similarly, Biswas et al. (2024) designed a hybrid explainable framework for detecting phishing and genuine URLs. This model used probability assessments score to evaluate cyber risks associated with phishing attacks. Pawlicki (2023) designed a network intrusion detection system that combined various deep learning models for performance evaluation. A decision tree for AI-driven cybersecurity services was suggested by Gerlach et al. (2022). Most of the above studies have used computational models such as Support Vector Machine (SVM), Decision Tree (DT), Classification and Regression Tree (CART), Convolutional Neural Networks (CNN), Deep Neural Networks (DNN) and Recurrent Neural Networks (RNN) for cyber threat detection. These models are effective at identifying patterns within static datasets but often lack the capability to adequately handle the dynamic, temporal dependencies inherent in network traffic data. In enterprise industrial network these limitations can pose significant hurdle in forecasting or recognizing evolving threats that develop over time. Additionally, these model lacks interpretation which is crucial for adjusting security measures and understanding underlying threat patterns. Recently, Khan et al. (2022) proposed an explainable autoencoder-based detection framework using CNN and RNN to identify cyber threats in industrial networks. This approach has shown robust performance, providing transparency in detecting malicious events and outperforming state-of-the-art methods. Similarly, Oseni et al. (2023) proposed an explainable CNN-based Intrusion Detection System (IDS). The proposed approach also used SHAP to interpret the decisions made by the model thereby enhancing transparency and allowing experts to understand and validate the IDS's effectiveness.

Based on the above discussion, we have formulated the following research questions (RQs) to guide our future investigations:

**RQ1:** *How effectively can GANs be applied to generate privacy-preserving synthetic data that accurately replicates real-world cyber attack patterns specifically within industrial networking environments, and how does this improve the training of threat detection systems in this domain?*

**RQ2:** *Given the limitations of traditional computational models in handling dynamic, temporal dependencies in network traffic data specific to cybersecurity, what novel approaches can be developed to enhance threat detection in industrial networks?*

**RQ3:** *How can the performance and interpretability of cyber threat detection models be improved through*

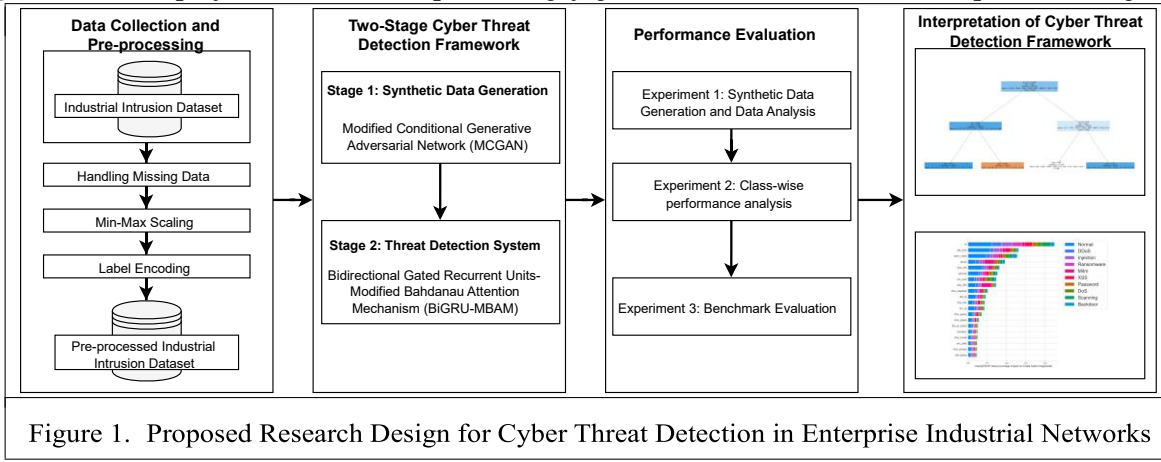

Figure 1. Proposed Research Design for Cyber Threat Detection in Enterprise Industrial Networks

*the interpretation techniques within the context of industrial network security?*

## Research Design

We propose a novel design to address the aforementioned research gaps. Figure 1 shows the proposed framework with four components: (1) data collection and preprocessing (2) two-stage cyber threat detection framework (3) performance evaluation, and (4) interpretation of cyber threat detection framework. We have explained each component in the subsections below.

### *Data collection and Pre-processing*

In the context of enterprise industrial networks, we have used two publicly available intrusion detection datasets for the experiment namely, ToN-IoT (Alsaedi et al. 2020) and Edge-IIoT (Ferrag et al. 2022) datasets. The ToN-IoT dataset is a new generation of Industry 4.0/IoT and IIoT dataset that includes various data types, such as system logs from IoT devices, network traffic, and physical information from IoT sensors. The dataset includes nine types of attacks, including Cross-Site Scripting (XSS), Distributed Denial of Service (DDoS), Denial of Service (DoS), Password Cracking, Reconnaissance, Man-In-The-Middle (MITM), Ransomware, Backdoors, and injection attacks. The Edge-IIoT dataset was generated on an IoT/IIoT testbed incorporating various devices, sensors, protocols, and configurations. It includes data from more than 10 types of IoT devices, such as temperature and humidity sensors, ultrasonic sensors, and heart rate monitors. The dataset categorizes 14 types of attacks into five main threats, including DoS/DDoS, information gathering, and malware attacks. The details of both datasets are mentioned in (Alsaedi et al. 2020) and (Ferrag et al. 2022).

The pre-processing of ToN-IoT and Edge-IIoT datasets includes three key steps: (1) handling missing data (2) label encoding, and (3) min-max scaling. The missing values in the dataset were filled using the median of that column. For instance, in the ToN-IoT dataset, missing values in the 'Idle Max' feature were filled using the median of that column due to its continuous nature. Similarly, in Edge-IIoT, categorical features like the 'HTTP.request.method', which includes HTTP methods such as GET and POST, were transformed into numerical codes using the label encoding method. Finally, features were scaled between 0 and 1 using the min-max scaling approach.

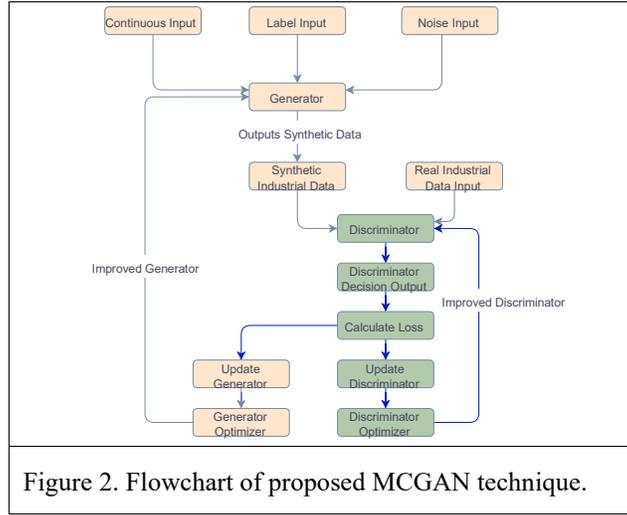

Figure 2. Flowchart of proposed MCGAN technique.

## Two-Stage Cyber Threat Detection Framework

### Stage 1: Synthetic Data Generation

In stage 1, the objective is to develop a method for generating synthetic network traffic data $\hat{y}_t$ that is representative of the complex, sequential nature of real network traffic data $y_t$ within enterprise industrial networks. This method must inherently support data anonymization to produce privacy-aware synthetic samples. Formally, we seek a function $G$ that maps from a latent space $z_t \in R^{d_z}$, possibly conditioned on auxiliary information $c_t$, to the space of network traffic data at time $t$, capturing the essential temporal dependencies and patterns for effective cybersecurity analysis and early threat detection. Additionally, the generated data $\hat{y}_t$ should preserve the statistical properties of $y_t$ while ensuring that any direct or indirect identifiers are anonymized.

In this context, the Generative Adversarial Network (GAN) emerges as a powerful solution. A GAN consists of two neural networks: the Generator (G) and the Discriminator (D), trained simultaneously through adversarial processes. The generator aims to produce data indistinguishable from real data, while the discriminator aims to distinguish between real and generated data. The objective function of a GAN is defined as: $\min_G \max_D V(D,G) = E_{x_t \sim p_{data}(x_t)}[\log D(x_t)] + E_{z_t \sim p_z(z_t)}[\log(1 - D(G(z_t)))]$; where $x_t$ is a sample from the real data distribution $p_{data}$ at time $t$, and $z_t$ is a random noise vector from distribution $p_z$, serving as the input to the generator at time $t$.

The Modified Conditional Generative Adversarial Network (MCGAN) is designed to generate synthetic network traffic data that reflects the sequential and complex nature of real-world network traffic, emphasizing privacy awareness and capturing temporal patterns. The inclusion of time $t$ alongside categorical labels $c_t$ and continuous features $f_t$ as conditional inputs allow for the generation of data that is not only contextually relevant but also temporally coherent. The generator ($G$) synthesizes data $\hat{y}_t$ at time $t$, using a latent noise vector $z_t$, and is conditioned on both categorical labels $c_t$ and continuous features $f_t$, reflecting the specific characteristics of network traffic at that moment: $G(z_t|c_t,f_t;\theta_G) \rightarrow \hat{y}_t$. The discriminator ($D$) assesses whether the data at time $t$ is real or synthetic, taking into account the same temporal and conditional information: $D(y_t,c_t,f_t;\theta_D)$ vs. $D(\hat{y}_t,c_t,f_t;\theta_D)$. The training objective for the MCGAN, capturing the adversarial dynamics between $G$ and $D$ while accommodating the conditional and temporal aspects, is given by: $\min_G \max_D V(D,G) = E_{y_t \sim p_{data}(y_t)}[\log D(y_t|c_t,f_t)] + E_{z_t \sim p_z(z_t)}[\log(1 - D(G(z_t|c_t,f_t)|c_t,f_t)]$. This formulation ensures the synthetic data generated at each time step $t$ is not only realistic and aligned with the specified conditions but also seamlessly integrates into the temporal sequence of network traffic, enhancing the model's utility for cybersecurity tasks such as threat detection and system training. The inclusion of $c_t$ and $f_t$ enriches the synthetic data, facilitating detailed simulations of network behaviors and cyber threats.

To further reinforce data privacy and anonymization, MCGAN incorporates a Conditional Feature Obfuscation (CFO) approach to ensure that the generated synthetic data does not depend on individual-specific data points, thereby minimizing the risk of disclosing sensitive information. In this framework, the original conditional features are obfuscated before they are used in the generator. Here, we can denote the original conditional features by $c_t$ and $f_t$. Then, we define an obfuscation function $O(c_t, f_t; \alpha)$, where $\alpha$ controls the level of obfuscation: $\tilde{c}_t = O(c_t; \alpha_c)$ and $\tilde{f}_t = O(f_t; \alpha_f)$ Here, $\tilde{c}_t$ and $\tilde{f}_t$ are the obfuscated features, and $\alpha_c$ and $\alpha_f$ are parameters controlling the obfuscation levels. The generator then takes $\tilde{c}_t$ and $\tilde{f}_t$ as inputs, alongside the latent vector $z_t$: $\hat{y}_t = G(z_t | \tilde{c}_t, \tilde{f}_t; \theta_G)$. To ensure that the generated data retains utility while preserving privacy, a privacy-preserving loss function $L_{\text{privacy}}$ is introduced. This loss function penalizes the generator if the generated data $\hat{y}_t$ is too closely aligned with the original features: $L_{\text{privacy}} = \lambda(\text{MSE}(G(z_t | \tilde{c}_t, \tilde{f}_t; \theta_G), G(z_t | c_t, f_t; \theta_G)))$. The total loss function for the generator is: $\min_G \max_D V(D,G) +$

$L_{\text{privacy}}$. This approach ensures that MCGAN generates synthetic data that is not only realistic and aligned with specified conditions but also robustly privacy-preserving, minimizing the risk of disclosing sensitive information. This approach promotes a safer framework for data sharing and collaborative efforts. Figure 2 explains the steps involved in designing the proposed MCGAN technique for synthetic data generation.

**Stage 2: Threat Detection System**

Although the MCGAN generates synthetic network traffic data, $\hat{y}_t$, it does not inherently decide when cybersecurity measures should be activated. This decision-making requires analyzing the synthetic data to understand when protective actions are needed. We employ a neural network architecture that integrates Bidirectional Gated Recurrent Units (BiGRU) and a Modified Bahdanau Attention Mechanism (MBAM) to process the synthetic data $\hat{y}_t$ and make informed decisions. The BiGRU layers, with forward $\overrightarrow{GRU}$ and backward $\overleftarrow{GRU}$, process the input data both in forward and reverse temporal directions, effectively capturing the sequential dependencies in the data. The forward pass of the GRU at each timestep $t$ is mathematically represented as $\overrightarrow{h}_t = \overrightarrow{GRU}(\hat{y}_t, \overrightarrow{h}_{t-1})$ and similarly, the backward pass is represented as $\overleftarrow{h}_t = \overleftarrow{GRU}(\hat{y}_t, \overleftarrow{h}_{t+1})$.

The output of the BiGRU at each timestep $t$, denoted as $h_t = [\overrightarrow{h}_t; \overleftarrow{h}_t]$, provides a comprehensive representation of the sequence, capturing both forward and backward dependencies within the synthetic network traffic data. This combination of forward and backward hidden states captures the complex temporal dynamics within network traffic, encompassing both past and future contextual insights within the sequence. However, while BiGRU layers effectively aggregate temporal information from both directions of the sequence, the challenge remains in discerning which parts of this rich temporal context are most relevant for identifying cybersecurity threats at any given moment. This is where the adaptation of the Bahdanau Attention Mechanism (BAM) becomes essential.

The standard Bahdanau Attention, originally conceived for natural language processing tasks, enables a model to dynamically focus on different segments of an input sequence to enhance sequential prediction performance. For instance, the BAM computes a context vector $c_t$ by focusing on relevant parts of the input sequence through a dynamically computed attention score: $e_{tj} = v_a^T \tanh(W_a h_t + U_a h_j + b_a)$, $\alpha_{tj} = \frac{\exp(e_{tj})}{\sum_{k=1}^{T} \exp(e_{tk})}$, $c_t = \sum_{j=1}^{T} \alpha_{tj} h_j$, where $e_{tj}$ calculates the alignment or energy score between the target hidden state $h_t$ and each source hidden state $h_j$ and $W_a, U_a, v_a, b_a$ denotes the initial weights and bias of the BAM. However, the direct application of this standard framework to the domain of cybersecurity, especially in analyzing synthetic network traffic data, encounters unique challenges. These challenges come from the distinct nature of network traffic data, which encompasses complex, non-linear temporal dependencies that are not typically present in NLP tasks.

To adapt the BAM for cybersecurity, especially in analyzing synthetic network traffic data, we introduce specific computational modifications that enhance its capability to identify complex patterns indicative of cyber threats. The weights $W_a$, $U_a$, and $v_a$ are tuned specifically using Adjust function to capture the temporal and spatial dependencies unique to network traffic data, which are not inherently sequential or linguistic and is computed as: $W_a' = \text{Adjust}(W_a)$, $U_a' = \text{Adjust}(U_a)$, $v_a' = \text{Adjust}(v_a)$, $b_a' = \text{Adjust}(b_a)$. The *Adjust* function is computed through an iterative optimization process, leveraging backpropagation with a categorical crossentropy loss function that quantifies the model's performance in accurately identifying cybersecurity threats. $W_a', U_a', v_a', b_a' = \arg\min_{W_a, U_a, v_a, b_a} L(y, \hat{y})$, where $W_a', U_a', v_a', b_a'$ are the tuned parameters of the attention mechanism, obtained through the optimization process. $L(y, \hat{y})$ denotes the loss function, which measures the discrepancy between the true labels $y$ and the predictions $\hat{y}$ made by the model using the current set of parameters. $y$ represents the true class labels indicating the presence or absence of a cybersecurity threat within the data. $\hat{y}$ denotes the model's predictions, computed based on the synthetic network traffic data processed through the model with the current parameters. The backpropagation algorithm is employed to compute the gradients of the loss function with respect to each parameter, allowing for the iterative adjustment of $W_a$, $U_a$, $v_a$, and $b_a$ in the direction that minimizes the loss. This optimization process ensures that the attention mechanism is increasingly refined to focus on the most relevant aspects of the network traffic data for threat detection. The adjustment is carried out across multiple epochs, with the model's performance on a validation set typically monitored to prevent overfitting and to ensure generalizability. $\Delta\theta = -\eta \nabla_\theta L$, where $\Delta\theta$ represents the change to be applied to each parameter ($\theta \in \{W_a, U_a, v_a, b_a\}$), $\eta$ is the learning rate, and $\nabla_\theta L$ is the gradient of the loss function with respect to $\theta$. Using these adjusted parameters, the computation $\cdot h_t + U_a' \cdot$ of the modified energy scores $e'_{tj}$, which now better reflect the cybersecurity context, is as follows: $e'_{tj} = v_a'^T \tanh(W_a' \tilde{h}_j + b_a' + C_{\text{context}})$, where $C_{\text{context}}$ adds an extra term to incorporate additional contextual features from the network traffic data, enhancing the model's ability to focus on relevant information. The attention weights and context vector are then updated to: $\alpha'_{t,j} = \frac{\exp(e'_{tj})}{\sum_{k=1}^{T} \exp(e'_{tk})}$, $c'_t = \sum_{j=1}^{T} \alpha'_{tj} \cdot \tilde{h}_j$. These equations highlight the novel adaptation of the Bahdanau Attention for cybersecurity, making it more adept at identifying significant patterns and anomalies in network traffic data. The $\tilde{h}_j$ represents an enhanced representation of input features, accounting for both the direct input and additional network context, ensuring a comprehensive analysis tailored to cybersecurity needs. Finally, to facilitate multi-class classification, a softmax layer is applied to the context vector $c'_t$, yielding the probability distribution over potential classes: $y_t = \text{softmax}(W_c \cdot c'_t + b_c)$, where $W_c$ and $b_c$ are the weight matrix and bias term for the classification layer, respectively, and $y_t$ represents the predicted probabilities for each class at time $t$.

### *Performance Evaluation*

In line with the computational design science paradigm (Yu et al. 2023), we thoroughly evaluated our proposed cyber threat detection artifact through a series of benchmark experiments (Ampel et al. 2024) against various machine learning and deep learning methods. The proposed framework was designed using Python 3.7 using Keras and TensorFlow frameworks. The framework uses NVIDIA A100 80GB PCIe GPU for development. The MCGAN architecture consists of a generator with 3 dense layers that has 256, 512, and 1024 neurons. The discriminator also has 3 layers with 1024, 512, and 256 neurons. The Adam optimizer, with a learning rate of 0.0002 and beta1 of 0.5, was used to train both networks. The BiGRU-MBAM architecture consists of 3 layers of BiGRU with 128 units with a learning rate of 0.0003. The MBAM attention scores were passed through a dense layer with 64 neurons, and final classification was performed using a softmax activation function to output probability distributions across the target classes. Hyperparameter tuning in proposed architecture and in the commonly used baselines of machine learning and deep learning methods were performed using a grid search approach.

In Experiment 1, we examined the synthetic data generated by MCGAN technique. This includes the performance comparison of GAN and MCGAN's generator and discriminator in terms of loss with increased epochs. Also, Experiment 1 includes analyzing the real and synthetic data distribution and density. In Experiment 2, we analyze the performance of the proposed cyber threat detection framework in terms of class-wise precision, detection rate, F1 score and False Alarm Rate (FAR). The formulas for each metrics is mentioned in (Kumar et al. 2022). For Experiment 3, we compare the performance against the most commonly used baselines of machine learning and deep learning methods. This includes comparing overall performance in terms of accuracy, precision, detection rate and F1 score. The experiment uses ToN-IoT (Alsaedi et al. 2020) and Edge-IIoT (Ferrag et al. 2022) datasets to benchmark the proposed cyber threat detection artifact, where each model uses stratified 10-fold cross-validation technique. The results section discusses the mean of each fold and includes one-tailed paired t-tests (i.e., *: p-values *: p<0.05, **: p<0.01, ***: p<0.001) to demonstrate the differences between the proposed approach and the baseline methods.

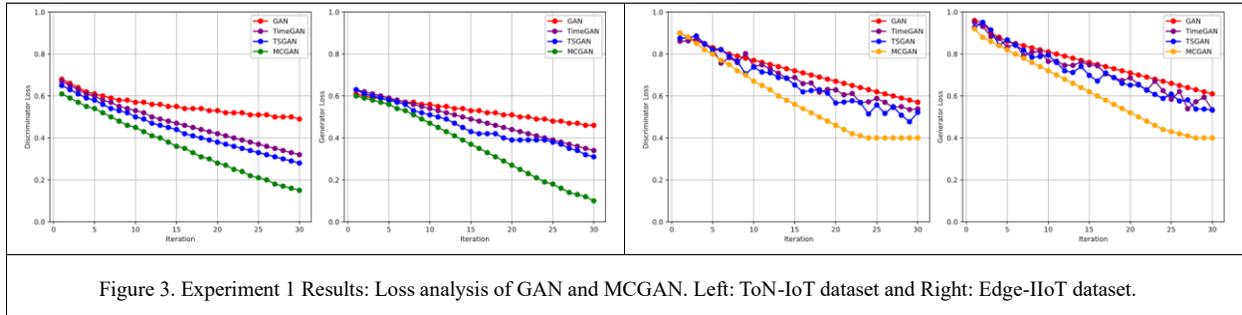

Figure 3. Experiment 1 Results: Loss analysis of GAN and MCGAN. Left: ToN-IoT dataset and Right: Edge-IIoT dataset.

### Interpretation of Cyber Threat Detection Framework

The third stage of our framework focuses on increasing the model's interpretability. Interpretability in machine learning refers to the capability to explain or to present in understandable terms to a human how the model makes its decisions. This stage involves the use of specific methods that make it possible to examine and understand the model's decision-making process, revealing the impact of input features on the model's predictions. 7

*Interpretation with SHAP*

Based on cooperative game theory, SHapley Additive exPlanations (SHAP) values offer a consistent and locally accurate method to attribute the prediction outcome to its input features. Given a model *f* and an input instance *x* with features $x_1, x_2, ..., x_n$, the contribution of a feature $x_j$ to the prediction is given by the SHAP value $\phi_j$, computed as: $\phi_j = \sum_{S \subseteq F \setminus \{j\}} \frac{|S|!(|F|-|S|-1)!}{|F|!} [f_x(S \cup \{j\}) - f_x(S)]$, where: *F* represents the set of all features, *S* is a subset of *F* excluding the feature *j*, $f_x(S)$ denotes the prediction when only features in *S* are observed, the term $[f_x(S \cup \{j\}) - f_x(S)]$ quantifies the marginal contribution of including feature *j* in the subset *S* (Lundberg and Lee 2017).

*Interpretation with Decision Tree Surrogate*

Decision Tree Classifiers are particularly useful for visualizing and understanding the feature interactions and thresholds that drive the predictions. Given a set of predictions from a complex model and the corresponding input features, a decision tree *T* can be trained as follows: (1) Input features *X* are flattened if necessary and used along with the predicted class labels *Y* to train the tree. (2) The goal is to minimize a measure of impurity *I* at each node, using metrics such as Gini impurity $I_G$ or entropy *H*, defined as: $I_G(p) = 1 - \sum_{i=1}^{C} p_i^2$, $H(p) = -\sum_{i=1}^{C} p_i \log_2(p_i)$, where $p_i$ is the proportion of samples of class *i* in the node, and *C* is the number of classes (Herbinger et al. 2023).

## Results and Discussion

### *Experiment 1:*

In Experiment 1, we assess the training dynamics of our MCGAN against GAN, TimeGAN and TSGAN using ToN-IoT and Edge-IIoT datasets. The primary focus is to evaluate how each models learn to generate synthetic data and improve their discriminative abilities over 30 epochs of training. Figure 3 shows this comparison using ToN-IoT and Edge-IIoT datasets, respectively. With ToN-IoT dataset, the generator loss for GAN starts at 0.68 and gradually decreases to 0.49, and discriminator loss shows same decreasing trend of 0.61 to 0.46. Same trend is seen with TimeGAN and TSGAN. In contrast, the proposed MCGAN shows better performance and the generator loss reduces from 0.61 to 0.15 and discriminator loss drops from 0.60 to 0.10. With Edge-IIoT dataset, the GAN generator loss decreases from 0.90 to 0.57, and the discriminator loss from 0.96 to 0.61. MCGAN again demonstrates superior performance with the generator loss reducing from 0.90 to a stable 0.40, and the discriminator loss decreasing from 0.92 to 0.40. The CFO framework in MCGAN allows the model to incorporate multiple conditional inputs (such as categorical labels and continuous features) while preserving privacy. This enables the generator to create synthetic data that is more representative of the real data distribution without overfitting to specific instances. This improves the losses of both generator and discriminator.

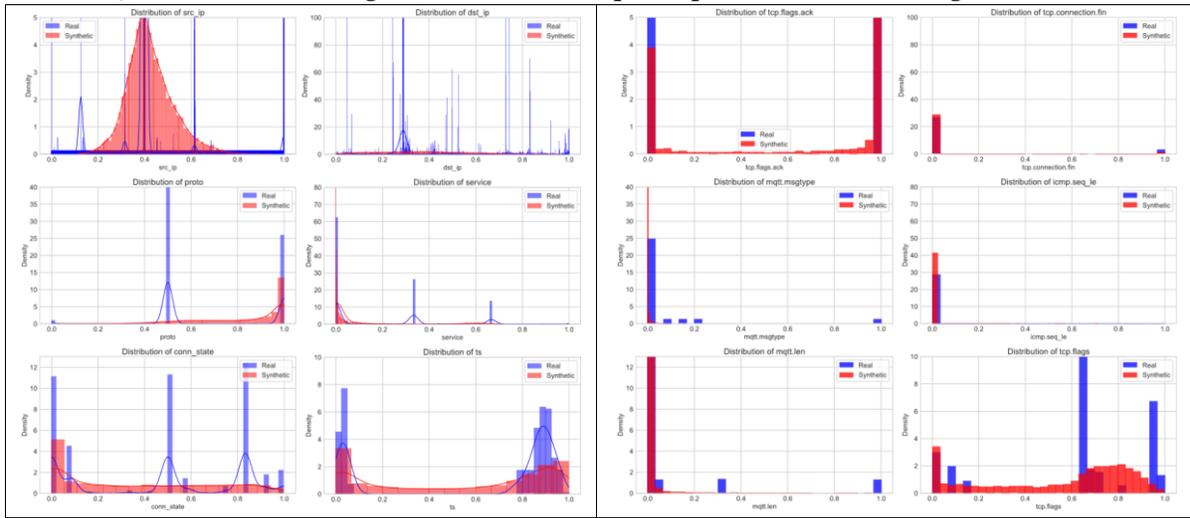

Figure 4. Experiment 1 Results: Analysis of real and synthetic network traffic data. Left: ToN-IoT and Right: Edge-IIoT dataset.

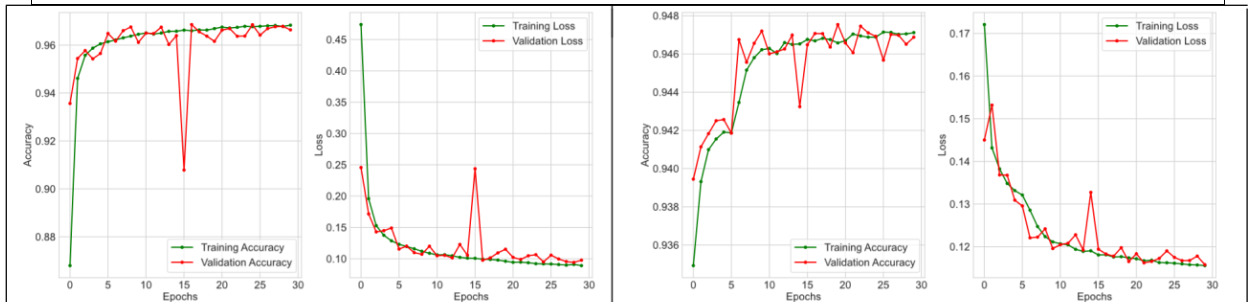

Figure 5. Experiment 2 Results: Training and Validation dynamics for BiGRU-MBAM. Left: ToN-IoT and Right: Edge-IIoT dataset.

Next, we have examined the data distributions and densities for key network features between real and synthetic datasets obtained from our proposed MCGAN technique. In Figure 4, the degree of overlap between the blue and red histograms for each feature shows how closely the synthetic data mimics the real data. It can be seen for both datasets that the proposed MCGAN was able to capture the essential statistical properties of the real data.

### *Experiment 2:*

In Experiment 2, we first analyze the training and validation dynamics for the BiGRU-MBAM based threat detection system over 30 epochs. The synthetic data was combined with actual dataset and then these anal-ysis was performed. Figure 5 shows this analysis for both dataset. The experiment is designed to evaluate the model's performance not only in terms of its ability to learn from the training data but also its generalizability as tested on a separate validation set. It can be noted from Figure 5 that the training and validation accuracy has increased for both datasets and training and validation loss has gradually decreased. This close convergence of training and validation indicate a well-tuned model that is likely to perform robustly on similar unseen data. However, we observed a temporary fluctuation in validation accuracy around epoch 15. This sudden change is due to a scheduled decrease in the learning rate. Such adjustments are part of our learning rate decay strategy, designed to refine the learning process as the model approaches convergence. The temporary drop in accuracy following this adjustment suggests a brief period where the model adapts to the finer gradients. However, it quickly returned to expected levels of accuracy, indicating that the model successfully adapted to the new learning rate. This adaptation underscores the effectiveness of our learning rate strategy, confirming its role in enhancing the model's ability to generalize from training to unseen data. The confusion matrix shown in Figure 6 illustrates the performance of the BiGRU-MBAM model in classifying network traffic data into various categories using confusion matrices for two distinct datasets: ToN-IoT and Edge-IIoT. In confusion matrix, each row of the matrix represents the instances in an actual class, while each column represents the instances in a predicted class. The diagonal cells represent the number of correct predictions made for each class, and the off-diagonal cells show where the model has made errors. The significant values along the diagonals of each matrix indicate that most of the attack instances across both datasets have been correctly identified by the model. However, off-diagonal cells highlight errors such as the false positives in SQL injection classifications, attributed due to the similar features of SQL injections and normal traffic. Similarly, Figure 7 illustrates the Receiver Operating Characteristic (ROC) curves for the BiGRU-MBAM applied to two different datasets: ToN-IoT and Edge-IIoT. The ROC curves for both datasets show Area Under the Curve (AUC) values close to 0.99, indicating high classification rate by the BiGRU-MBAM model.

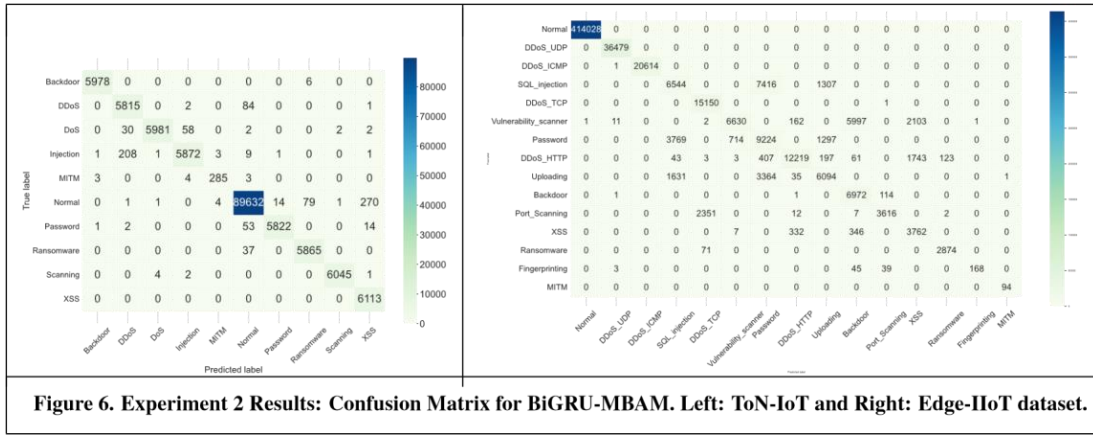

Figure 6. Experiment 2 Results: Confusion Matrix for BiGRU-MBAM. Left: ToN-IoT and Right: Edge-IIoT dataset.

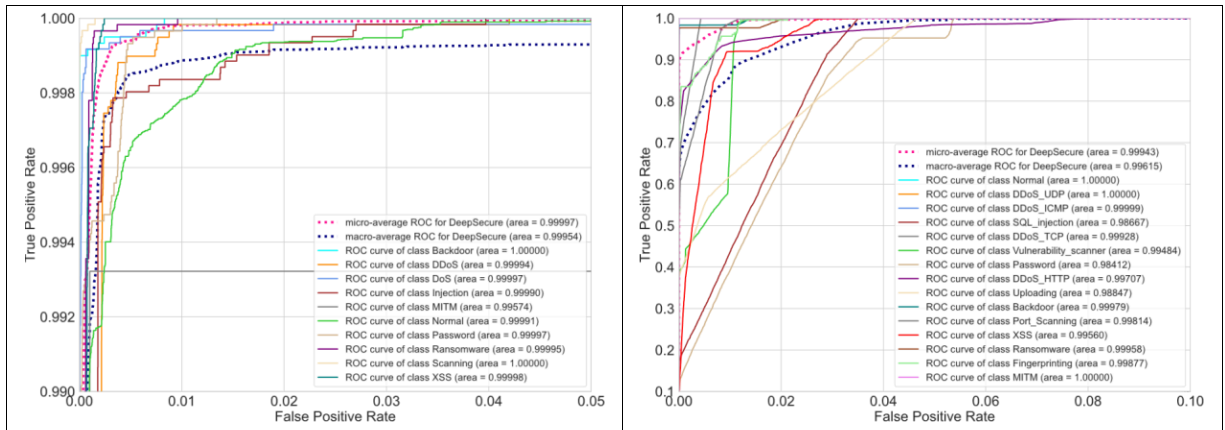

Figure 7. Experiment 2 Results: ROC Curve for BiGRU-MBAM. Left: ToN-IoT and Right: Edge-IIoT dataset.

| Metrics | Backdoor | DDoS | DoS | Injection | MITM | Normal | Password | Ransomware | Scanning | XSS |
|---|---|---|---|---|---|---|---|---|---|---|
| PR | 99.91 | 96.02 | 99.89 | 98.88 | 97.60 | 99.79 | 99.74 | 98.57 | 99.95 | 95.48 |
| DR | 99.89 | 98.52 | 98.45 | 96.32 | 96.61 | 99.58 | 98.81 | 99.37 | 99.88 | 100.00 |
| F1 | 99.99 | 97.25 | 99.17 | 97.59 | 97.10 | 99.68 | 99.27 | 98.97 | 99.91 | 97.69 |
| FAR | 0.00003 | 0.00182 | 0.00004 | 0.00049 | 0.00005 | 0.00389 | 0.00011 | 0.00064 | 0.00002 | 0.00218 |

Table 1. Experiment 2 Results: Class-wise performance analysis of BiGRU-MBAM on different metrics using ToN-IoT dataset.

| Metrics | Normal | DDoS_UDP | DDoS_ICMP | SQL_Injection | DDoS_TCP | VS | Password | DDoS_HTTP | Uploading | Backdoor | Port_Scanning | XSS | Ransom | Fingerprint | MITM |
|---|---|---|---|---|---|---|---|---|---|---|---|---|---|---|---|
| PR | 99.99 | 99.95 | 100.00 | 54.59 | 86.19 | 90.15 | 45.19 | 95.75 | 68.51 | 51.92 | 95.91 | 49.44 | 95.83 | 99.40 | 98.94 |
| DR | 100.00 | 100.00 | 99.99 | 42.86 | 99.99 | 44.48 | 61.48 | 82.57 | 54.78 | 98.36 | 60.39 | 84.60 | 97.59 | 65.88 | 100.00 |
| F1 | 100.00 | 99.98 | 100.00 | 48.02 | 92.58 | 59.57 | 52.09 | 88.67 | 60.88 | 67.97 | 74.11 | 62.41 | 96.70 | 79.25 | 99.47 |
| FAR | 0.00001 | 0.00003 | 0.00000 | 0.00967 | 0.00431 | 0.00129 | 0.01986 | 0.00096 | 0.00494 | 0.01130 | 0.00027 | 0.00670 | 0.00022 | 0.00000 | 0.00000 |

Table 2. Experiment 2 Results: Class-wise performance analysis of BiGRU-MBAM on different metrics using Edge-IIoT dataset.

| Model | Accuracy | Precision | Detection Rate | F1 Score |
|---|---|---|---|---|
| RF | 68.29* | 65.20* | 64.17** | 60.19** |
| NB | 71.67** | 68.11** | 67.73** | 66.36** |
| SVM | 77.11** | 69.05** | 68.02** | 70.59** |
| DNN | 88.68*** | 67.98** | 64.48*** | 61.73*** |
| LSTM | 91.24*** | 72.71*** | 70.28*** | 68.36*** |
| GRU | 94.23*** | 77.53*** | 77.23*** | 76.93*** |
| BiLSTM | 91.86*** | 72.20*** | 71.83*** | 71.34*** |
| BiGRU | 95.86*** | 95.60*** | 95.25*** | 95.41*** |
| BiGRU-MBAM | 99.35 | 99.82 | 98.74 | 98.65 |

Table 3. Experiment 3 Results: Benchmark Evaluations using ToN-IoT dataset.

| Model | Accuracy | Precision | Detection Rate | F1 Score |
|---|---|---|---|---|
| RF | 76.21*** | 29.22*** | 41.01*** | 45.19*** |
| NB | 79.97*** | 30.11*** | 41.15*** | 47.95*** |
| SVM | 81.24*** | 31.18*** | 41.22*** | 48.44*** |
| DNN | 89.51* | 34.24* | 41.84* | 35.94* |
| LSTM | 91.44*** | 54.79*** | 53.56*** | 50.33*** |
| GRU | 89.41** | 39.21** | 45.46** | 38.27** |
| BiLSTM | 93.64*** | 79.44*** | 66.61*** | 66.92*** |
| BiGRU | 93.53*** | 73.14*** | 64.59*** | 65.69*** |
| BiGRU-MBAM | 94.17 | 85.81 | 79.53 | 78.77 |

Table 4. Experiment 3 Results: Benchmark Evaluations using Edge-IIoT dataset.

Table 1 and Table 2 in Experiment 2 present the class-wise performance analysis of the BiGRU-MBAM model on the ToN-IoT and Edge-IIoT datasets, respectively. The proposed model demonstrates high PR and DR across most attack types, indicating its efficiency in accurately identifying specific threats. For example, the PR for Backdoor, DDoS, and Scanning attacks in ToN-IoT dataset is close to 100%. On the other hand, Table 2 shows a variation in performance, with some attack types like SQL Injection and Password exhibiting lower precision, which may point to challenges in detecting complex or less frequent attack patterns. Despite variations in precision, the DR for attacks remain high, indicating the model's strong capability to recognize attacks once they are accurately identified and the FAR are close to 0%.

**Experiment 3:**

In Experiment 3, we evaluated the performance of the proposed BiGRU-MBAM against the baseline machine learning and deep learning techniques. Table 3 and Table 4 shows this comparison against Random Forest (RF), Naive Bayes (NB), Support Vector Machine (SVM), Deep Neural Network (DNN), Long ShortTerm Memory (LSTM), Gated Recurrent Unit (GRU), Bidirectional LSTM (BiLSTM), and Bidirectional GRU (BiGRU) using ToN-IoT and Edge-IIoT datasets, respectively. From Table 3 and Table 4, we can see that the RNN-based models have achieved better performance than the classical ML methods. The BiGRU-MBAM shows superior performance compared to other models and achieves an accuracy of 99.35%, precision of 99.82%, recall of 98.74%, and an F1-score of 98.65% using ToN-IoT dataset. Similarly, the performance of BiGRU-MBAM using Edge-IIoT dataset is high and with an accuracy of 94.17%, precision of 85.81%, recall of 79.53%, and an F1-score of 78.77%. These findings show that including more contextual factors (continuous input, label input, and noise input) in the MCGAN considerably increases its ability to generate realistic synthetic industrial network data. Furthermore, including the BiGRU-MBAM for threat detection significantly improves performance. This collaborative approach improves the system's ability to handle complex sequential input while simultaneously improving its decision-making process.

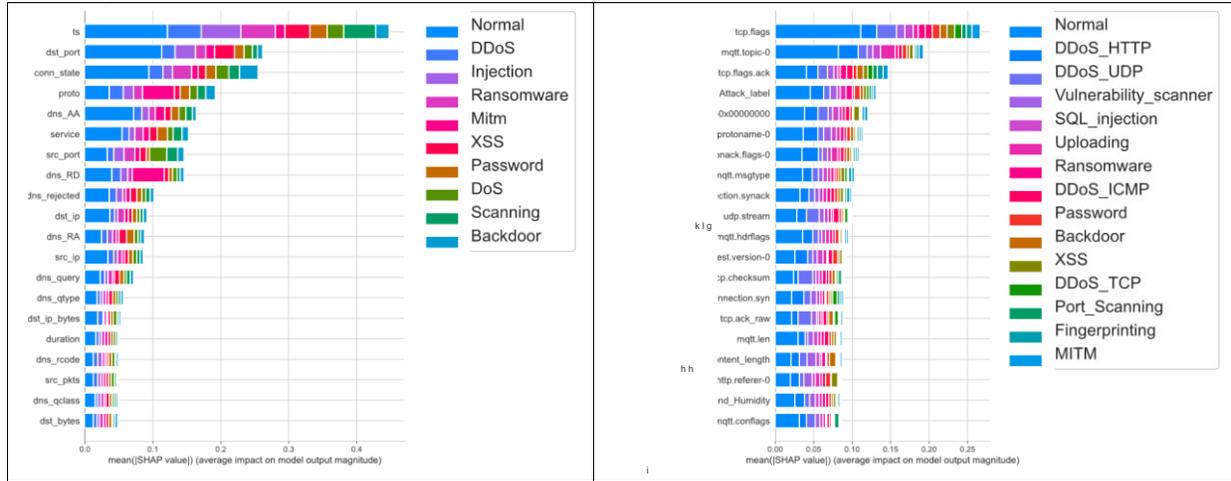

Figure 8. Interpretation using SHAP. Left: ToN-IoT dataset and Right: Edge-IIoT dataset.

### *Interpretation of proposed cyber threat detection framework*

The SHAP plot in Figure 8 visually demonstrates the importance of various network traffic features that the proposed cyber threat detection framework prioritizes for threat detection using ToN-IoT and Edge IIoT datasets. Both datasets highlight the significance of protocol-specific features and connection details. In ToN-IoT, features like proto and conn_state were pivotal, while in Edge-IIoT, tcp.flags, tcp.flags.ack, and MQTT-related features such as mqtt.topic-0 and mqtt.msgtype are crucial. This indicates a consistent focus across both datasets on the critical role of protocol behaviors in identifying cybersecurity threats. On the other hand, features that detail communication patterns, such as tcp.connection.synack in Edge-IIoT and dst_port in ToN-IoT, similarly demonstrate high impact, underscoring the model's reliance on specific aspects of network communications to make predictions. The emphasis on these features aligns with their relevance in network security, where the nature of traffic—defined by when it occurs, where it is directed, how it is connected, and over which protocol—can significantly determine its threat level. However, the Edge-IIoT dataset exhibits a more pronounced emphasis on IIoT-specific protocols (MQTT), which is expected given the IIoT-centric nature of edge computing environments. Features like mqtt.conack.flags-0x00000000 and mqtt.protoname-0 highlight the detailed understanding required to effectively monitor and secure enterprise industrial communications. The insights from SHAP value analysis across both datasets underscore the need for cybersecurity strategies that are tailored to the specific network architectures and traffic patterns prevalent in IIoT and traditional network environments.

The decision tree visualization generated from the surrogate model provides an insightful interpretation of the complex decision-making process inherent in our proposed cyber threat detection framework. Figure 9 illustrates the decision tree derived from the proposed model predictions, showcasing how initial splits based on "src_ip" and "ts" guide the classification process using ToN-IoT dataset. Each node describes the Gini impurity, sample size, and class distribution, demonstrating the model's reliance on crucial network properties to distinguish between normal operations and potential cyber threats. Similarly, Figure 10 depicts the decision tree resulting from the surrogate model analysis using Edge-IIoT dataset. Each node indicates a split based on features like "icmp.checksum", "udp.stream", and "Attack_label", and is quantified by metrics such as Gini impurity, sample size, and class distribution across various threat categories. This provides clear insight into which features are deemed most informative by the model, aiding stakeholders in refining both the feature engineering and model architecture for improved performance in cybersecurity threat detection within enterprise industrial environments.

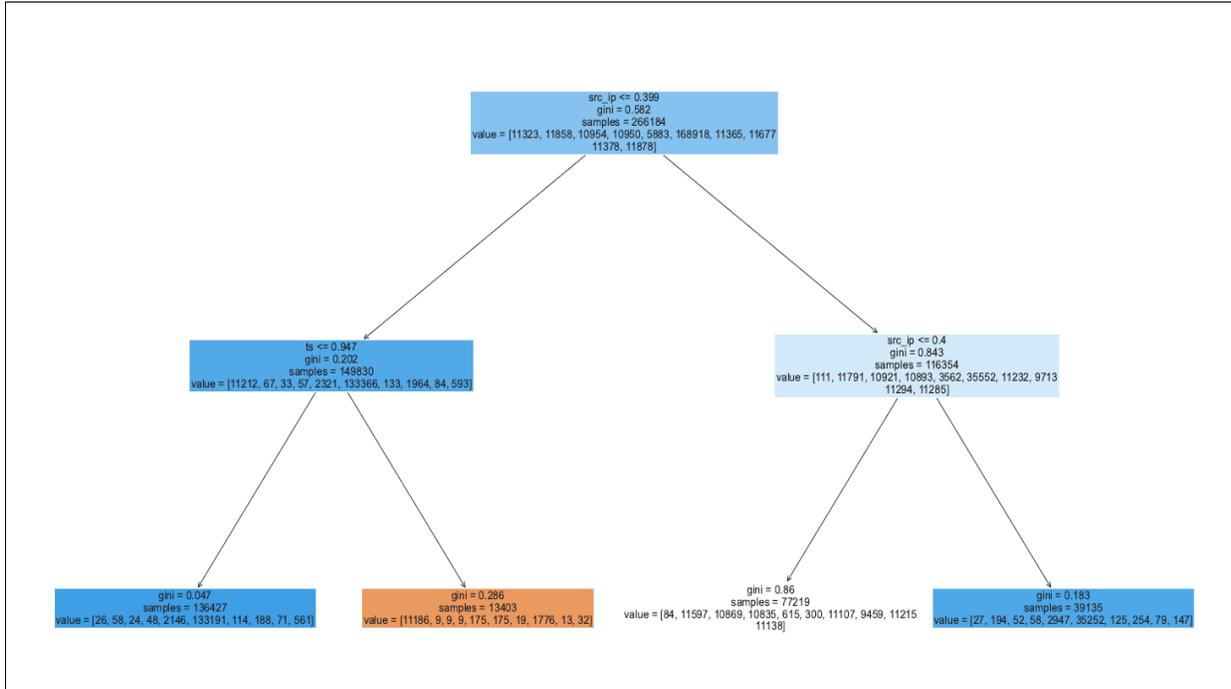

Figure 9. Interpreting the decisions through a Surrogate Decision Tree Model on ToN-IoT dataset

## Discussion: Contributions to the IS Knowledge Base and Practical Implications

Following the principles of computational design science as outlined by (Yu et al. 2023), we focused our research on cybersecurity applications, specifically within enterprise industrial networks (Singh et al. 2023). We thoroughly investigated and assessed potential solutions to develop an innovative IT artifact tailored to the specific needs of our research inquiry. We will detail these contributions and their corresponding practical implications in the subsequent subsections.

### *Contribution to the IS Knowledge Base*

IS researchers assert that innovative computational IT artifacts must enhance the IS knowledge base with prescriptive knowledge (Hevner et al. 2008). This contribution can manifest as new constructs, models, methods, instantiations, or design theories (Ampel et al. 2024). Furthermore, while a computational methodology might be developed and evaluated within a specific functional context, its principles can often be generalized and applied across multiple domains (Samtani et al. 2022). We believe the design principles offered in this study fit these guidelines. Specifically, our proposed interpretable cyber threat detection system contributes two novel design principles to the IS knowledge base:

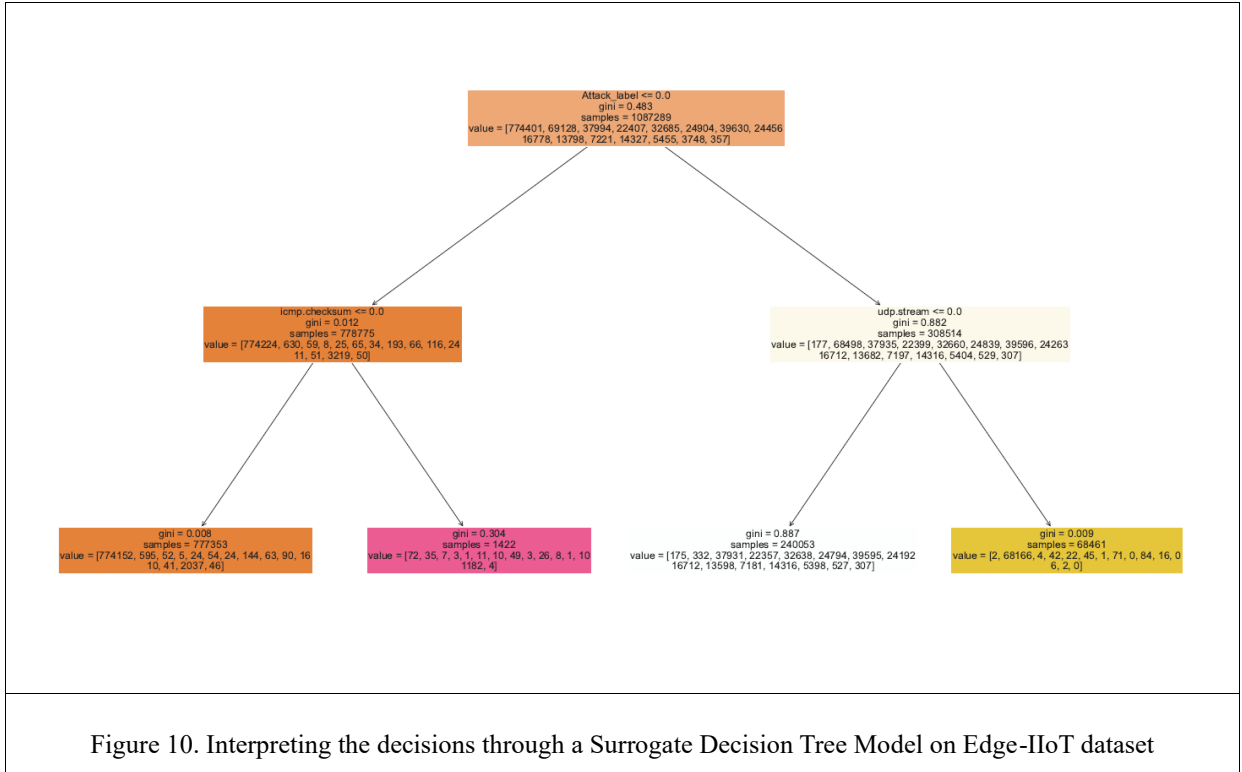

Figure 10. Interpreting the decisions through a Surrogate Decision Tree Model on Edge-IIoT dataset

(1) a modified conditional generative adversarial network (MCGAN) to generate synthetic network traffic data that reflects the sequential and complex nature of real-world network traffic, emphasizing privacy awareness and capturing temporal patterns.

(2) a novel threat detection mechanism based on a bidirectional gated recurrent units and a modified bahdanau attention mechanism (BiGRU-MBAM) is designed. This system utilizes bidirectional processing of sequential data to capture temporal relationships and dependencies that are pivotal for recognizing anomalous patterns. The integration of the MBAM allows the model to focus selectively on segments of the input data that are most indicative of cybersecurity threats, effectively weighing the importance of each timestep's contribution to the overall decision-making process. By dynamically adjusting its focus and combining insights from both past and future contexts within the data, BiGRU-MBAM conducts a thorough analysis of network activities. This enables it to detect complex threats with a higher level of precision than traditional models.

Although we developed the design principles 1 and 2 for cybersecurity analytics in the context of enterprise systems, but they could be applied to other research inquiries in different bodies of IS research. The proposed MCGAN and BiGRU-MBAM can be beneficial in several key areas that are central to IS research:

(1) *Health Informatics:* In health informatics, the availability of diverse and extensive datasets is crucial for training effective machine learning models, especially in tasks like disease prediction or patient outcome forecasting (Rasouli Panah et al. 2023). MCGAN can generate synthetic medical data that mimics real patient data, which is particularly beneficial where privacy concerns or rarity of certain medical conditions limit the availability of actual data. This synthetic data can be used to augment existing datasets, providing a richer, more varied dataset for model training without compromising patient privacy. The BiGRU-MBAM excels in analyzing time-series data common in health informatics, such as patient monitoring data or sequential entry of electronic health records (EHRs). This model can be integral to systems used for continuous patient monitoring, where it can help in real-time decision-making and early warning systems. For example, analyzing trends in heart rate or blood sugar levels to predict and alert to potential health crises before they occur.

(2) *SentimentAnalysisandSocialMediaMonitoring:* By applying MCGAN and BiGRU-MBAM, researchers can more effectively analyze social media data to gauge public sentiment and monitor social trends (Dhar and Bose 2022). Specifically, MCGAN can be useful for training sentiment analysis models when the available real data is insufficient or biased. Moreover, MCGAN-generated data can be utilized to expand the training dataset for sentiment analysis models, especially in scenarios where certain sentiments or topics are underrepresented in the available datasets. The BiGRU-MBAM is particularly effective in social media monitoring due to its ability to process sequential data, such as time-stamped social media posts. Moreover, it can be deployed in real-time social media monitoring systems to track public sentiment trends, monitor the spread of information, and identify sudden shifts in public opinion or emerging social issues.

### *Practical Implications*

The proposed cyber threat detection system can provide significant benefits to different professionals within the enterprise industrial network context.

(1) *Cybersecurity Professionals:* The design principle 1 can assist Security Operations Center (SOC) teams to generate and utilize synthetic network traffic data without the risk of exposing sensitive information. Moreover, they can also improve their AI models for detecting different types of attacks without the need to collect and store massive amounts of real network traffic. Thus, this approach can enable them to develop and refine threat detection and response strategies effectively, tailored to the specific needs of enterprise industrial networks. The design principle 2 can also assist SOC and Incident Response Managers (IRM) to detect and respond to threats with greater precision. This capability is essential for minimizing response times and reducing the impact of security breaches on industrial operations.

(2) *Enterprise Network Administrators:* The synthetic data generated from design principle 1 can be used by enterprise network administrators to safely test and develop network security policies without risking exposure of actual sensitive network traffic. Moreover, the synthetic data generation supports data protection regulations, such as GDPR or HIPAA in healthcare, by minimizing the need to use real data that could potentially expose personal or critical information. Administrators can develop an effective tool for risk assessment and management strategies by using the generated synthetic data. Moreover, the design principle 2 can be used to develop a proactive threat monitoring system (i.e., can capture temporal dependencies in network traffic data) that enables the detection of potential security issues before they escalate into breaches.

## Conclusion and Future Directions

Detecting cyber threats in enterprise industrial networks requires a detailed understanding of both the technological and operational landscape. Networks in industrial settings are not only vast and complex but also critical to the functioning of the enterprises they support. Ensuring their security involves monitoring and analyzing network traffic for anomalies that could indicate malicious activities. In this study, we aimed to develop a novel approach that systematically addresses the challenges of cyber threat detection in enterprise industrial networks. Our methodology was structured in two main stages. The first stage involved creating synthetic network traffic data using a Modified Conditional Generative Adversarial Network (MCGAN). This step was crucial for overcoming the limitations associated with the scarcity of real-world cyber threat data and the privacy concerns related to using sensitive information. In the second stage, we implemented a neural network architecture that combines Bidirectional Gated Recurrent Units (BiGRU) with a Modified Bahdanau Attention Mechanism. This configuration was specifically designed to capture the dynamic and complex nature of network traffic data effectively. Further, to ensure the practical applicability and transparency of our approach, we also integrated Shapley Additive Explanations (SHAP) and a decision tree surrogate model. Through benchmark evaluation on two open-source datasets, this framework demonstrated its effectiveness in identifying cyber threats with exceptional precision over some commonly used models. Future research will focus on enhancing the capabilities of this model to support real-time detection and on investigating its applicability across various sectors within the IS field.

## Acknowledgements

This work was supported by the Research Council of Finland with CHIST-ERA, grant agreement no - 359790, Di4SPDS-Distributed Intelligence for Enhancing Security and Privacy of Decentralized and Distributed Systems.

## References

Alsaedi, A., Moustafa, N., Tari, Z., Mahmood, A., and Anwar, A. (2020). TON_IoT telemetry dataset: A new generation dataset of IoT and IIoT for data-driven intrusion detection systems. *Ieee Access* (8), pp. 165130–165150. DOI: 10.1109/ACCESS.2020.3022862.

Ampel, B. M., Samtani, S., Zhu, H., Chen, H., and Nunamaker Jr, J. F. (2024). Improving Threat Mitigation Through a Cybersecurity Risk Management Framework: A Computational Design Science Approach. *Journal of Management Information Systems* (41:1), pp. 236–265. DOI: https://doi.org/10.1080/07421222.2023.2301178.

Anande, T. and Leeson, M. (2023). Synthetic Network Traffic Data Generation and Classification of Advanced Persistent Threat Samples: A Case Study with GANs and XGBoost. In: *International Conference on Deep Learning Theory and Applications,* Springer, pp. 1–18. DOI: https://doi.org/10.1007/978-3-031-39059-3_1.

Beverungen, D., Buijs, J. C., Becker, J., Di Ciccio, C., van der Aalst, W. M., Bartelheimer, C., vom Brocke, J., Comuzzi, M., Kraume, K., Leopold, H., et al. (2021). Seven paradoxes of business process management in a hyper-connected world. *Business & Information Systems Engineering* (63), pp. 145–156. DOI: https: //doi.org/10.1007/s12599-020-00646-z.

Biswas, B., Mukhopadhyay, A., Kumar, A., and Delen, D. (2024). A hybrid framework using explainable AI (XAI) in cyber-risk management for defence and recovery against phishing attacks. *Decision Support Systems* (177), p. 114102. DOI: https://doi.org/10.1016/j.dss.2023.114102.

De, S., Bermudez-Edo, M., Xu, H., and Cai, Z. (2022). Deep Generative Models in the Industrial Internet of Things: A Survey. *IEEE Transactions on Industrial Informatics* (18:9), pp. 5728–5737. DOI: 10.1109/TII.2022.3155656.

Dhar, S. and Bose, I. (2022). Walking on air or hopping mad? Understanding the impact of emotions, sentiments and reactions on ratings in online customer reviews of mobile apps. *Decision Support Systems* (162), p. 113769. DOI: https://doi.org/10.1016/j.dss.2022.113769.

Ferrag, M. A., Friha, O., Hamouda, D., Maglaras, L., and Janicke, H. (2022). Edge-IIoTset: A new comprehensive realistic cyber security dataset of IoT and IIoT applications for centralized and federated learning. *IEEE Access* (10), pp. 40281–40306. DOI: 10.1109/ACCESS.2022.3165809.

Gerlach, J., Werth, O., and Breitner, M. H. (2022). Artificial Intelligence for Cybersecurity: Towards Taxonomybased Archetypes and Decision Support. In: *ICIS,* DOI: https://aisel.aisnet.org/icis2022/security/ security/10.

Herbinger, J., Dandl, S., Ewald, F. K., Loibl, S., and Casalicchio, G. (2023). Leveraging Model-Based Trees as Interpretable Surrogate Models for Model Distillation. In: *European Conference on Artificial Intelligence,* Springer, pp. 232–249. DOI: https://doi.org/10.1007/978-3-031-50396-2_13.

Hevner, A. R., March, S. T., Park, J., and Ram, S. (2008). Design science in information systems research. *Management Information Systems Quarterly* (28:1), p. 6. DOI: https://doi.org/10.2307/25148625.

Johnson, M., Albizri, A., Harfouche, A., and Fosso-Wamba, S. (2022). Integrating human knowledge into artificial intelligence for complex and ill-structured problems: Informed artificial intelligence. *International Journal of Information Management* (64), p. 102479. DOI: https://doi.org/10.1016/j.ijinfomgt. 2022.102479.

Kayhan, V. O., Agrawal, M., and Shivendu, S. (2023). Cyber threat detection: Unsupervised hunting of anomalous commands (UHAC). *Decision Support Systems* (168), p. 113928. DOI: https://doi.org/10.1016/j.dss.2023.113928.

Khan, I. A., Moustafa, N., Pi, D., Sallam, K. M., Zomaya, A. Y., and Li, B. (2022). A New Explainable Deep Learning Framework for Cyber Threat Discovery in Industrial IoT Networks. *IEEE Internet of Things Journal* (9:13), pp. 11604–11613. DOI: 10.1109/JIOT.2021.3130156.

Kisselburgh, L. and Beever, J. (2022). The ethics of privacy in research and design: Principles, practices, and potential. In: *Modern socio-technical perspectives on privacy,* Springer International Publishing Cham, pp. 395–426. DOI: https://doi.org/10.1007/978-3-030-82786-1_17.

Kumar, P., Kumar, R., Gupta, G. P., Tripathi, R., and Srivastava, G. (2022). P2tif: A blockchain and deep learning framework for privacy-preserved threat intelligence in industrial iot. *IEEE Transactions on Industrial Informatics* (18:9), pp. 6358–6367. DOI: 10.1109/TII.2022.3142030.

Lim, W., Yong, K. S. C., Lau, B. T., and Tan, C. C. L. (2024). Future of generative adversarial networks (GAN) for anomaly detection in network security: A review. *Computers Security* (139), p. 103733. DOI: https://doi.org/10.1016/j.cose.2024.103733.

Lin, Z., Shi, Y., and Xue, Z. (2022). Idsgan: Generative adversarial networks for attack generation against intrusion detection. In: *Pacific-asia conference on knowledge discovery and data mining,* Springer, pp. 79–91. DOI: https://doi.org/10.1007/978-3-031-05981-0_7.

Lundberg, S. M. and Lee, S.-I. (2017). A unified approach to interpreting model predictions. *Advances in neural information processing systems* (30). DOI: 10.5555/3295222.3295230.

Mozo, A., González-Prieto, Á., Pastor, A., Gómez-Canaval, S., and Talavera, E. (2022). Synthetic flow-based cryptomining attack generation through Generative Adversarial Networks. *Scientificreports* (12:1), p. 2091. DOI: https://doi.org/10.1038/s41598-022-06057-2.

Nedeljkovic, D. and Jakovljevic, Z. (2022). CNN based method for the development of cyber-attacks detection algorithms in industrial control systems. *Computers & Security* (114), p. 102585. DOI: https://doi.org/10.1016/j.cose.2021.102585.

Oseni, A., Moustafa, N., Creech, G., Sohrabi, N., Strelzoff, A., Tari, Z., and Linkov, I. (2023). An Explainable Deep Learning Framework for Resilient Intrusion Detection in IoT-Enabled Transportation Networks. *IEEE Transactions on Intelligent Transportation Systems* (24:1), pp. 1000–1014. DOI: 10.1109/TITS.2022.3188671.

Pawlicki, M. (2023). Strengths And Weaknesses of Deep, Convolutional and Recurrent Neural Networks in Network Intrusion Detection Deployments. *ISD 2023 Proceedings* (). DOI: https://doi.org/10.62036/ISD.2023.54.

Quach, S., Thaichon, P., Martin, K. D., Weaven, S., and Palmatier, R. W. (2022). Digital technologies: Tensions in privacy and data. *Journal of the Academy of Marketing Science* (50:6), pp. 1299–1323. DOI: https://doi.org/10.1007/s11747-022-00845-y.

Rasouli Panah, H., Madanian, S., and Yu, J. (2023). Integration of AI and Big Data Analysis with Public Health Systems for Infectious Disease Outbreak Detection. *ACIS 2023 Proceedings. 41.* (). DOI: https://aisel.aisnet.org/acis2023/41.

Research, P. M. (2024). Industrial Networking Solutions Market Size Worth USD 150.68 Billion By 2032. Online; accessed 1-March-2023.

Samtani, S., Chai, Y., and Chen, H. (2022). Linking Exploits from the Dark Web to Known Vulnerabilities for Proactive Cyber Threat Intelligence: An Attention-based Deep Structured Semantic Model. *MIS quarterly* (46:2). DOI: https://doi.org/10.25300/MISQ/2022/15392.

Singh, N., Krishnaswamy, V., and Zhang, J. Z. (2023). Intellectual structure of cybersecurity research in enterprise information systems. *Enterprise Information Systems* (17:6), p. 2025545. DOI: https://doi.org/10.1080/17517575.2022.2025545.

Unhelkar, B. and Arntzen, A. A. (2020). A framework for intelligent collaborative enterprise systems. Concepts, opportunities and challenges. *Scandinavian Journal of Information Systems* (32:2), p. 6. DOI: https://aisel.aisnet.org/sjis/vol32/iss2/6.

Yang, Z., Li, Y., and Zhou, G. (2023). Ts-gan: Time-series gan for sensor-based health data augmentation. *ACM Transactions on Computing for Healthcare* (4:2), pp. 1–21. DOI: 10.1145/3583593.

Yoon, J., Jarrett, D., and Van der Schaar, M. (2019). Time-series generative adversarial networks. *Advances in neural information processing systems* (32).

Yu, S., Chai, Y., Samtani, S., Liu, H., and Chen, H. (2023). Motion Sensor–Based Fall Prevention for Senior Care: A Hidden Markov Model with Generative Adversarial Network Approach. *Information Systems Research* (). DOI: https://doi.org/10.1287/isre.2023.1203.

Yu Chung Wang, W., Pauleen, D., and Taskin, N. (2022). Enterprise systems, emerging technologies, and the data-driven knowledge organisation. *Knowledge Management Research & Practice* (20:1), pp. 1–13. DOI: https://doi.org/10.1080/14778238.2022.2039571.